\def\rn{}
\def\nn#1 #2{#2. #1}			       
\def\nnn#1 #2 #3{#2. #3. #1}		       
\def\nnnn#1 #2 #3 #4{#2. #3. #4 #1}	       
\def\nnnnn#1 #2 #3 #4 #5{#2. #3. #4 #5. #1}    
\def\dualand{ and\hbox{ }}				
\def\multiand{, and\hbox{ }}				
\def\rf#1;#2;#3;#4;#5 {{\frenchspacing\par\rn#1, #3 {\bf #4}, #5 (#2). \par}}
\def\rfbook#1;#2;#3;#4;#5 {{\frenchspacing\par\rn#1, {\it #3} (#5, #4, #2).\par}}
\def\rfproc#1;#2;#3;#4;#5;#6 {{\frenchspacing\par\rn#1, in {\it #3}, ed. #4 (#6, #5, #2).\par}}
\def\rfprep#1;#2;#3 {{\par\frenchspacing\rn#1 #2, #3\par}}

\def\seconds{{\rm s}}

\def\cm{{\rm cm}}
\def\nm{{\rm nm}}
\def\um{\mu{\rm m}}

\def\expec#1{\langle#1\rangle}

\def\nothing{\noindent\centerline{$\hbox{\,}$}}

\def\etal{{\frenchspacing\it et al.}}
\def\ie{{\frenchspacing\it i.e.}}
\def\eg{{\frenchspacing\it e.g.}}
\def\etc{{\frenchspacing\it etc.}}

\def\crr{\cr\noalign{\vskip 4pt}}

\def\beq#1{\begin{equation}\label{#1}}
\def\eeq{\end{equation}}
\def\beqa#1{\begin{eqnarray}\label{#1}}
\def\eeqa{\end{eqnarray}}
\def\eq#1{equation~(\ref{#1})}
\def\Eq#1{Equation~(\ref{#1})}

\def\fig#1{Figure~\arabic{#1}}
\def\Fig#1{Figure~\arabic{#1}}
\def\mycaption#1#2{\footnotesize{\bf FIG.~\arabic{#1}.} #2}

\def\sec#1{Section~\ref{#1}}

\def\simlt{\lesssim}

\def\ed{\end{document}}

\def\a{{\bf a}}

\def\p{{\bf p}}
\def\q{{\bf q}}
\def\r{{\bf r}}
\def\x{{\bf x}}
\def\y{{\bf y}}

\def\bzero{{\bf 0}}
\def\ah{\widehat{\a}}
\def\ph{\widehat{p}}

\def\yh{\widehat{\y}}
\def\I{{\bf I}}
\def\F{{\bf F}}
\def\M{{\bf M}}
\def\Sig{{\bf\Sigma}}

\def\Hint{H_{\rm int}}
\def\Hsubj{H_{\rm s}}
\def\Hobj{H_{\rm o}}
\def\Henv{H_{\rm e}}
\def\Hso{H_{\rm so}}
\def\Hoe{H_{\rm oe}}
\def\Hse{H_{\rm se}}
\def\rsubj{\rho_{\rm s}}
\def\robj{\rho_{\rm o}}

\def\rso{\rho_{\rm so}}

\def\tdiss{\tau_{\rm diss}}

\def\tdec{\tau_{\rm dec}}
\def\tdyn{\tau_{\rm dyn}}
\def\tcoll{\tau_{\rm coll}}
\def\leff{\lambda}
\def\water{{\rm H}_2{\rm O}}
\def\Naplus{{\rm Na}^+}
\def\Kplus{{\rm K}^+}
\def\tr{\hbox{tr}\,}
\def\ith{i^{th}}

\def\first{1^{st}}
\def\second{2^{nd}}
\def\tensormult{\otimes}

\def\blankface{\,\ddot{\raisebox{-2pt}{-}}\,}
\def\smileface{\raisebox{-2pt}{$\ddot\smile$}}
\def\frownface{\raisebox{-2pt}{$\ddot\frown$}}

\def\ket#1{|#1\rangle}

\def\noobs{\ket{\blankface}}
\def\upobs{\ket{\smileface}}
\def\downobs{\ket{\frownface}}

\def\up{\ket{\!\!\uparrow}}
\def\down{\ket{\!\!\downarrow}}
\def\noup{\ket{\blankface\!\!\uparrow}}
\def\nodown{\ket{\blankface\!\!\downarrow}}
\def\upup{\ket{\smileface\!\!\uparrow}}
\def\updown{\ket{\smileface\!\!\downarrow}}
\def\downup{\ket{\frownface\!\!\uparrow}}
\def\downdown{\ket{\frownface\!\!\downarrow}}

\def\bra#1{\langle #1|}

\def\noobsbra{\bra{\blankface}}
\def\upobsbra{\bra{\smileface}}
\def\downobsbra{\bra{\frownface}}

\def\upbra{\bra{\uparrow\!\!}}
\def\downbra{\bra{\downarrow\!\!}}
\def\noupbra{\bra{\blankface\!\!\uparrow\!\!}}
\def\nodownbra{\bra{\blankface\!\!\downarrow\!\!}}
\def\upupbra{\bra{\smileface\!\!\uparrow\!\!}}

\def\downdownbra{\bra{\frownface\!\!\downarrow}}

\documentstyle[pra,twocolumn,aps,epsf]{revtex}
\begin{document}

\newcounter{TimescaleFig}
\newcounter{TrinityFig}
\newcounter{NeuronFig}
\newcounter{ChessFig}
\newcounter{ChessFig2}
\setcounter{TimescaleFig}{1}
\setcounter{TrinityFig}{2}
\setcounter{NeuronFig}{3}
\setcounter{ChessFig}{4}
\setcounter{ChessFig2}{5}




\title{The Importance of Quantum Decoherence in Brain Processes}

\author{Max Tegmark}

\address{Institute for Advanced Study, Olden Lane,
Princeton, NJ 08540; max@ias.edu\\
Dept.~of Physics, Univ.~of Pennsylvania, Philadelphia, PA 19104\\
(Submitted to Phys. Rev. E July 2 1999, accepted October 25)
}

\maketitle

\begin{abstract}
Based on a calculation of neural decoherence rates, we argue that
that the degrees of freedom of the human brain that relate to 
cognitive processes
should be thought of as a classical rather than quantum system,
\ie, that there is nothing fundamentally wrong with 
the current classical approach to neural network simulations.
We find that the decoherence timescales ($\sim 10^{-13}-10^{-20}$ seconds) 
are typically much shorter than the relevant dynamical 
timescales ($\sim 10^{-3}-10^{-1}$ seconds),
both for regular
neuron firing and for kink-like polarization excitations
in microtubules. This conclusion disagrees
with suggestions by Penrose and others
that the brain acts as a quantum computer, and
that quantum coherence is related to consciousness in a fundamental way.
\end{abstract}


%
%

\section{Introduction}

In most current mainstream biophysics research on 
cognitive processes, 
the brain is modeled as a neural network obeying 
classical physics.
In contrast, Penrose \cite{Penrose89,Penrose97},
and others
have argued that quantum mechanics 
may play an essential role, 
and that successful brain simulations can only be 
performed with a quantum computer.
The main purpose of this paper is to address this issue 
with quantitative decoherence calculations.

The field of artificial neural networks 
(for an introduction, see, \eg, \cite{Amit,Mezard,Harvey})
is currently booming,
driven by a broad range of applications and improved computing resources.
Although the popular neurological models come in various levels of abstraction,
none involve effects of quantum coherence in any fundamental way.
Encouraged by successes in modeling memory, learning, 
visual processing, \etc \cite{Eeckman,McMillen99}, many 
workers in the field have boldly conjectured that
a sufficiently complex neural network could in principle 
perform all cognitive processes that we associate with 
consciousness.

On the other hand, many authors have argued that
consciousness can only be understood as a quantum effect.
For instance, 
Wigner \cite{Wigner62} suggested that consciousness was linked to 
the quantum measurement problem\footnote{
Interestingly, Wigner changed his mind and gave up this idea \cite{Wigner95}
after he became aware in of the first paper on decoherence
in 1970 \cite{Zeh70}.
}, and 
this idea has been greatly elaborated by Stapp \cite{StappBook}.
There have been numerous suggestions that 
consciousness is a macroquantum effect, involving
superconductivity \cite{Walker},
superfluidity \cite{Domash},
electromagnetic fields \cite{Stapp93},
Bose condensation \cite{Marshall,Zohar},
superflourescence \cite{Rosu}
or
some other mechanism \cite{Umezawa,Vitiello95}.
Perhaps the most concrete one is that
of Penrose \cite{Penrose97}, proposing that 
this takes place in microtubules, the 
ubiquitous hollow cylinders that among other things
help cells maintain their shapes.
It has been argued that microtubules can process information
like a cellular automaton \cite{Hameroff}, and Penrose
suggests that they operate as a quantum computer.
This idea has been further elaborated 
employing string theory methods
\cite{Nano95,Mavro95a,Mavro95b,Mavro95c,Mavro98a,Mavro98b,Mavro99}.

The make-or-break issue for all these quantum models is whether
the relevant degrees of freedom of the brain can be 
sufficiently isolated to retain their quantum coherence,
and opinions are divided.
For instance, Stapp has argued that interaction with
the environment is probably small enough to be unimportant for
certain neural processes \cite{Stapp91}, whereas 
Zeh \cite{Zeh81}, 
Zurek \cite{Zurek91},
Scott \cite{Scott96},
Hawking \cite{Hawking97} and
Hepp \cite{Hepp99}
have conjectured that environment-induced coherence will 
rapidly destroy macrosuperpositions in the brain.
It is therefore timely to try to settle the issue
with detailed calculations of the relevant decoherence rates.
This is the purpose of the present work.

The rest of this paper is organized as follows.
In \sec{SubsystemSec}, we briefly review the 
open system quantum mechanics
necessary for our calculations, and 
introduce a decomposition into three subsystems
to place the problem in its proper context.
In \sec{CalculationSec}, we evaluate decoherence rates 
both for neuron firing and for the microtubule processes
proposed by Penrose {\etal}, relegating some technical 
details to the Appendix.
We conclude in \sec{DiscussionSec} by discussing
the implications of our results, both for modeling 
cognitive brain processes and for incorporating them into
a quantum-mechanical treatment of the 
rest of the world.

\section{Systems and subsystems}
\label{SubsystemSec}

In this section, we review those aspects of 
quantum mechanics for open systems that are needed
for our calculations, and 
introduce a classification scheme and a 
subsystem decomposition 
to place the problem at hand in its appropriate context.

\subsection{Notation}

Let us first briefly review the quantum mechanics of subsystems.
The state of an arbitrary quantum system is described by its density
matrix $\rho$, which left in isolation will evolve in time according to the
Schr\"odinger equation
\beq{HeisenbergEq}
\dot\rho = -i[H,\rho]/\hbar.
\eeq
It is often useful to view a system as composed of two subsystems,
so that some of the degrees of freedom correspond to the 
$\first$ and the rest to the $\second$.
The state of subsystem $i$ is described by the
reduced density matrix $\rho_i$ obtained by tracing (marginalizing)
over the degrees of freedom of the other:
$\rho_1\equiv\tr_2\rho$, $\rho_2\equiv\tr_1\rho$.
Let us decompose the Hamiltonian as 
\beq{HintEq}
H = H_1 + H_2 + H_{int},
\eeq
where the operator $H_1$ affects only 
the $\first$ subsystem and $H_2$ 
affects only the $\second$ subsystem.
The interaction Hamiltonian $H_{int}$ is 
the remaining nonseparable part, defined
as $H_{int}\equiv H - H_1 - H_2$, so such a decomposition is
always possible, although it is generally only useful 
if $H_{int}$ is in some sense small.

If $\Hint=0$, \ie, if there is no interaction between the two subsystems,
then it is easy to show that 
$\dot\rho_i = -i[H_i,\rho_i]/\hbar$, $i=1,2$,
that is, 
we can treat each subsystem as if the rest of the Universe did not exist,
ignoring any correlations with the other subsystem that may have been
present in the full non-separable density matrix $\rho$. 
It is of course this property that makes density matrices so 
useful in the first place, and that led von Neumann 
to invent them \cite{Neumann}: 
the full system is assumed to 
obey \eq{HeisenbergEq}
simply because its interactions with the rest of the Universe
are negligible.

\subsection{Fluctuation, dissipation, communication and decoherence}

In practice, the interaction $\Hint$ between subsystems is usually
not zero. This has a number of qualitatively different effects:
\begin{enumerate}
\item Fluctuation
\item Dissipation
\item Communication
\item Decoherence
\end{enumerate}
The first two involve transfer of energy between the subsystems,
whereas the last two involve exchange of information.
The first three occur in classical physics as well - only the
last one is a purely quantum-mechanical phenomenon.

For example, consider a tiny colloid grain (subsystem 1) in a jar of water
(subsystem 2). Collisions with water molecules will cause 
{\bf fluctuations} in the center-of-mass position of 
the colloid (brownian motion). If its initial velocity
is high, {\bf dissipation} (friction) will slow it down to a mean speed
corresponding to thermal equilibrium with the water.
The dissipation timescale $\tdiss$, defined as the
time it would take to lose half of the initial excess energy,
will in this case be of order $\tcoll\times (M/m)$, where
$\tcoll$ is the mean-free time between collisions,
$M$ the colloid mass $M$ and $m$ is the mass of a water molecule.
We will define {\bf communication} as exchange of information.
The information that the two subsystems have about each other,
measured in bits, 
is 
\beq{InfoDefEq}
I_{12}\equiv S_1 + S_2 - S,
\eeq
where $S_i\equiv -\tr_i\rho_i\log\rho_i$ is the entropy of 
the $\ith$ subsystem, $S\equiv -\tr\rho\log\rho$ is the entropy of 
the total system, and the logarithms are base 2.
If this mutual information is zero, then the states of the two systems 
are uncorrelated and independent, with the density matrix of 
the separable form $\rho=\rho_1\tensormult\rho_2$.
If the subsystems start out independent, any interaction will at least initially
increase the subsystem entropies $S_i$, 
thereby increasing the mutual information, since the 
entropy $S$ of the total system always remains constant.

This apparent entropy increase of subsystems, which is related to the
arrow of time and the $\second$ law of of thermodynamics 
\cite{ZehTime}, occurs also in classical physics.
However, quantum mechanics produces a qualitatively new effect as well,
known as {\bf decoherence} \cite{Zeh70,Zurek81,Zurek82},
suppressing off-diagonal elements in the reduced density matrices 
$\rho_i$.
This effect destroys the ability to observe 
long-range quantum superpositions within the subsystems, 
and is now rather well-understood and uncontroversial 
\cite{Zurek91,Zurek84,Peres,Pearle,Gallis,Unruh} 
-- the interested reader is referred to \cite{Omnes97} 
and a recent book on decoherence \cite{ZehBook} for details.
For instance, if our colloid was initially in a superposition of 
two locations separated by a centimeter, this macrosuperposition
would for all practical purposes be destroyed by the first 
collision with a water molecule, \ie, on a timescale
$\tdec$ of order $\tcoll$, with
the quantum superposition surviving only on scales
below the de de Broigle wavelength of the water molecules
\cite{JZ85,collapse}.\footnote{Decoherence 
picks out a preferred basis in the quantum-mechanical Hilbert space,
termed the ``pointer basis'' by Zurek \cite{Zurek81}, in which superpositions
are rapidly destroyed and classical behavior is approached.
This normally includes the position basis, which is why we never
experience superpositions of objects in macroscopically different positions.
Decoherence is quite generic. 
Although it has been claimed that this preferred basis consists of
the maximal set of commuting observables that also commute with
$\Hint$ (the ``microstable basis'' of Omnes \cite{Omnes97}), 
this is in fact merely a sufficient condition, not a necessary one.
If $[\Hint,x]=0$ for some observable $x$ but $[\Hint,p]\ne 0$
for its conjugate $p$, then the interaction will indeed cause 
decoherence for $x$ as advertised. But this will happen
even if $[\Hint,x]\ne 0$  --- all that matters is that
$[\Hint,p]\ne 0$, \ie, that the interaction Hamiltonian
contains (``measures'') $x$.
}
This means that $\tdiss/\tdec\sim M/m$ in our example, 
\ie, that decoherence is much faster than dissipation 
for macroscopic objects, 
and this qualitative result has been shown to hold 
quite generally as well (see \cite{Omnes97} and references therein).
Loosely speaking, this is because each microscopic particle 
that scatters off of the subsystem carries away only a tiny fraction 
$m/M$ of the total momentum, but essentially all of
the necessary information.

\begin{figure}[tb] 
\centerline{\epsfxsize=3.4in\epsffile{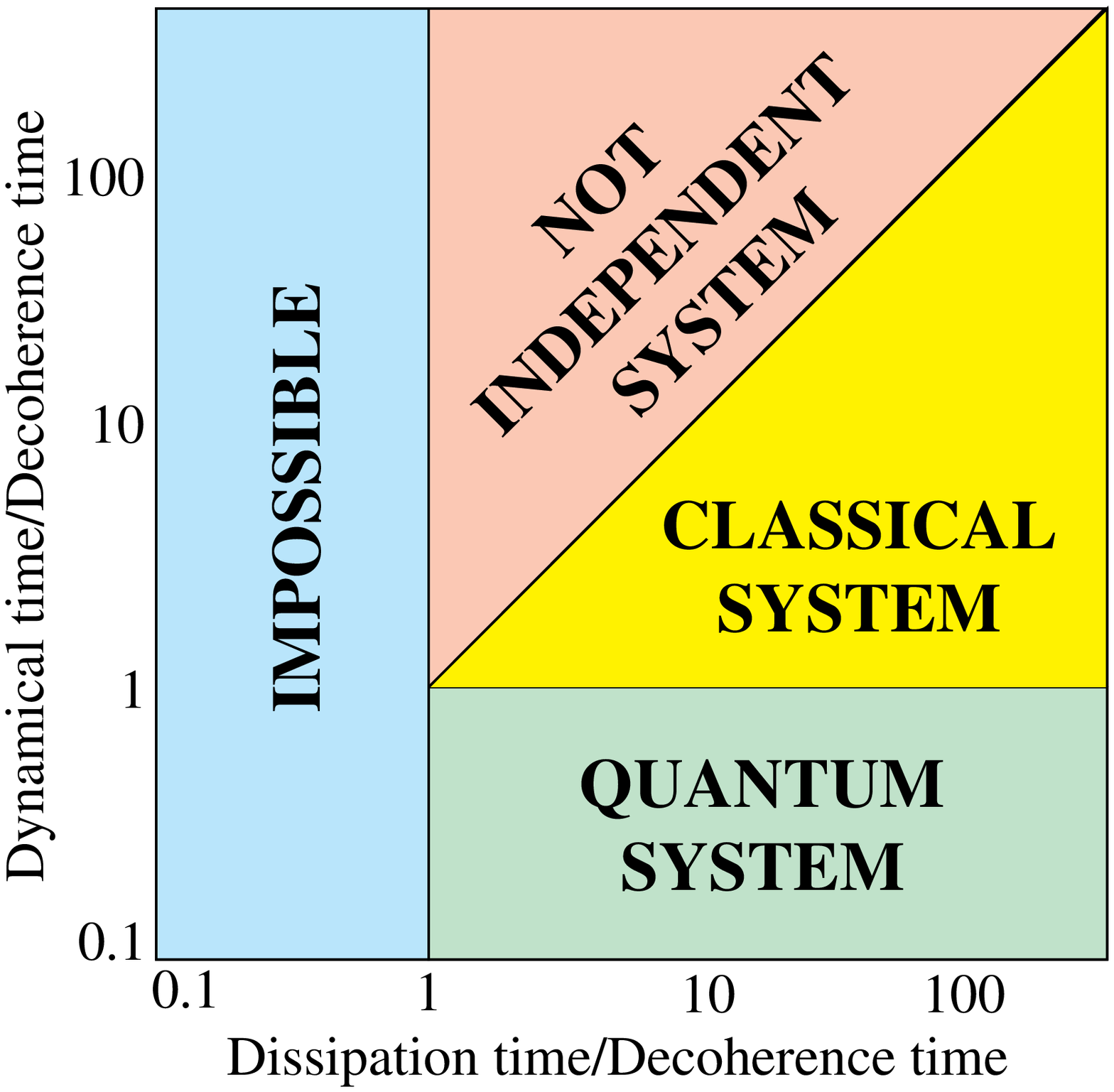}}
\vskip0.3cm
\mycaption{TimescaleFig}{The qualitative 
behavior of a subsystem depends on 
the timescales for dynamics, dissipation and decoherence.
This classification is by necessity quite crude, so 
the boundaries should not be thought of as sharp.
}
\end{figure}

\subsection{Classification of systems}

Let us define the {\it dynamical timescale} $\tdyn$ of
a subsystem as that which is characteristic of its internal dynamics.
For a planetary system or an atom,
$\tdyn$ would be the orbital frequency.

The qualitative behavior of a system depends on the ratio of 
these timescales, as illustrated in \fig{TimescaleFig}.
If $\tdyn\ll\tdec$, we are are dealing with a true quantum system, since
its superpositions can persist long enough to be dynamically important.
If $\tdyn\gg\tdiss$, it is hardly meaningful to view it as an independent
system at all, since its internal forces are so week that they 
are dwarfed by the effects of the surroundings.
In the intermediate case where 
$\tdec\ll\tdyn\simlt\tdiss$, we have a familiar classical system.  

The relation between $\tdec$ and $\tdiss$ depends only on 
the {\it form} of $\Hint$, whereas the question of whether 
$\tdyn$ falls between these values depends on the
{\it normalization} of $\Hint$ in \eq{HintEq}.
Since $\tdec\sim\tdiss$ for microscopic (atom-sized) systems
and $\tdec\ll\tdiss$ for macroscopic ones, 
\fig{TimescaleFig} shows that whereas macroscopic
systems can behave quantum-mechanically, 
microscopic ones can never behave classically.

\subsection{Three systems: subject, object and environment}

Most discussions of quantum statistical mechanics
split the Universe into two subsystems \cite{Feynman72}:
the object under consideration 
and everything else
(referred to as the {\it environment}).
Since our purpose is to model the observer, we need
to include a third subsystem as well, the {\it subject}.
As illustrated in \fig{TrinityFig}, we therefore decompose the
total system into three subsystems: 
\begin{itemize}
\item The {\bf subject} consists of the degrees 
of freedom associated with the 
subjective perceptions of the observer.
This does not include any other degrees of freedom associated
with the brain or other parts of the body.

\item The {\bf object} consists of the degrees of freedom 
that the observer is interested in studying, \eg, the pointer
position on a measurement apparatus.

\item The {\bf environment} consists of everything else,
\ie, all the degrees of freedom that the observer is 
not paying attention to.
By definition, these are the degrees of freedom that we 
always perform a partial trace over.

\end{itemize}
\begin{figure}[tb] 
\nothing\vskip-2cm
\centerline{\epsfxsize=3.45in\epsffile{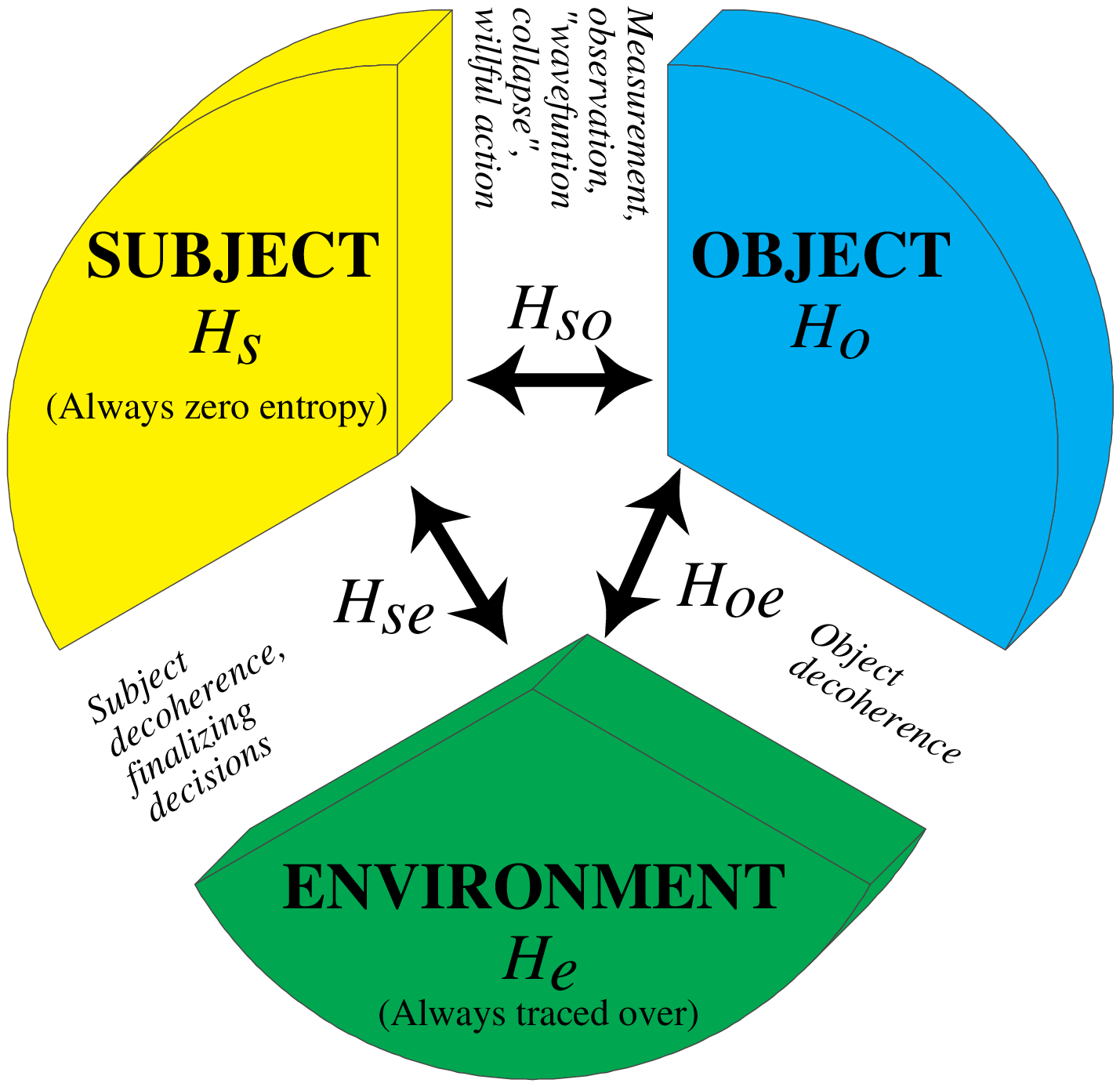}}
\vskip0.5cm
\mycaption{TrinityFig}{An observer 
can always decompose the world into three subsystems:
the degrees of freedom corresponding to her subjective
perceptions (the subject),
the degrees of freedom being studied (the object), 
and everything else (the environment).
As indicated, the subsystem Hamiltonians 
$\Hsubj$, $\Hobj$, $\Henv$ and 
the interaction Hamiltonians
$\Hso$, $\Hoe$, $\Hse$ can
cause qualitatively very different effects,
which is why it is often useful to study them separately.
This paper focuses on the interaction $\Hse$.
}
\end{figure}
Note that the first two definitions are very restrictive.
Whereas the subject would include the entire body of the observer 
in the common way of speaking, 
only very few degrees of freedom qualify as 
our subject or object.
For instance, 
if a physicist is observing a Stern-Gerlach apparatus, the
vast majority of the $\sim 10^{28}$ degrees of freedom in the
the observer and apparatus are counted as 
environment, not as subject or object.

The term ``perception'' is used in a broad sense in item 1, 
including thoughts, emotions and any other attributes of the 
subjectively perceived state of the observer.

The practical usefulness in this decomposition lies in that 
one can often neglect everything except the object
and its internal dynamics (given by $\Hobj$) to first order,
using simple prescriptions to correct for the interactions
with the subject and the environment.
The effects of both $\Hso$ and
$\Hoe$ have been extensively studied in the literature.
$\Hso$ involves quantum measurement, 
and gives rise to the usual interpretation of the diagonal elements
of the object density matrix as probabilities.
$\Hoe$ produces decoherence, selecting a preferred basis
and making the object act classically
if the conditions in \fig{TimescaleFig} are met.

In contrast, $\Hse$, which causes decoherence directly in 
the subject system, has received relatively little 
attention. It is the focus of the present paper,
and the next section is devoted to quantitative 
calculations of decoherence in brain processes,
aimed at determining whether the subject system should be 
classified as classical or quantum in the sense of 
\fig{TimescaleFig}.
We will return to \fig{TrinityFig} and a more detailed discussion
of its various subsystem interactions in \sec{DiscussionSec}.

\section{Decoherence rates}
\label{CalculationSec}

In this section, we will make quantitative estimates
of decoherence rates for neurological processes.
We first analyze the process of neuron firing, 
widely assumed to be central to cognitive processes.
We also analyze electrical excitations in microtubules,
which Penrose and others have suggested may be relevant to
conscious thought.

\subsection{Neuron firing}

Neurons (see \fig{NeuronFig}) are one of the key
building blocks of the brain's
information processing system.
It is widely believed that the complex network of 
$\sim 10^{11}$ neurons with 
their nonlinear synaptic couplings
is in some way linked to our subjective perceptions,
\ie, to the subject degrees of freedom.
If this picture is correct, then 
if $\Hsubj$ or $\Hso$ puts the subject into 
a superposition of two distinct mental states, 
some neurons will be in a superposition of firing and not firing. 
How fast does such a superposition of a firing and non-firing neuron 
decohere? 

Let us consider this process in more detail.
For introductory reviews of neuron dynamics, the reader
is referred to, \eg, \cite{Katz,Schade,Cairns-Smith}.
Like virtually all animal cells, neurons have ATP driven 
pumps in their membranes which push sodium ions out of
the cell into the surrounding fluids and 
potassium ions the other way.
The former process is slightly more efficient, so 
the neuron contains a slight excess of negative charge
in its ``resting'' state, corresponding to a potential 
difference $U_0\approx -0.07\,$V across the axon membrane 
(``axolemma'').
There is an inherent instability in the system, however.
If the potential becomes substantially less negative, then
voltage-gated sodium channels in the axon membrane open up, 
allowing $\Naplus$ ions to come gushing in. This makes 
the potential still less negative, causes still more 
opening, {\it etc}. This chain reaction, ``firing'', propagates down the
axon at a speed of up to 100 m/s, changing the
potential difference to a value $U_1$ that is typically of order
$+0.03\,$V \cite{Schade}.

The axon quickly recovers. After less than $\sim 1\,$ms, the 
sodium channels close regardless of the voltage, and 
large potassium channels (also voltage gated, but with a time delay)
open up allowing $\Kplus$ ions 
to flow out and restore the resting potential $U_0$.
The ATP driven pumps quickly restore the $\Naplus$ and
$\Kplus$ concentrations to their initial values, making 
the neuron ready to fire again if triggered.
Fast neurons can fire over $10^3$ times per second.

\begin{figure}[tb] 
\centerline{\epsfxsize=3.4in\epsffile{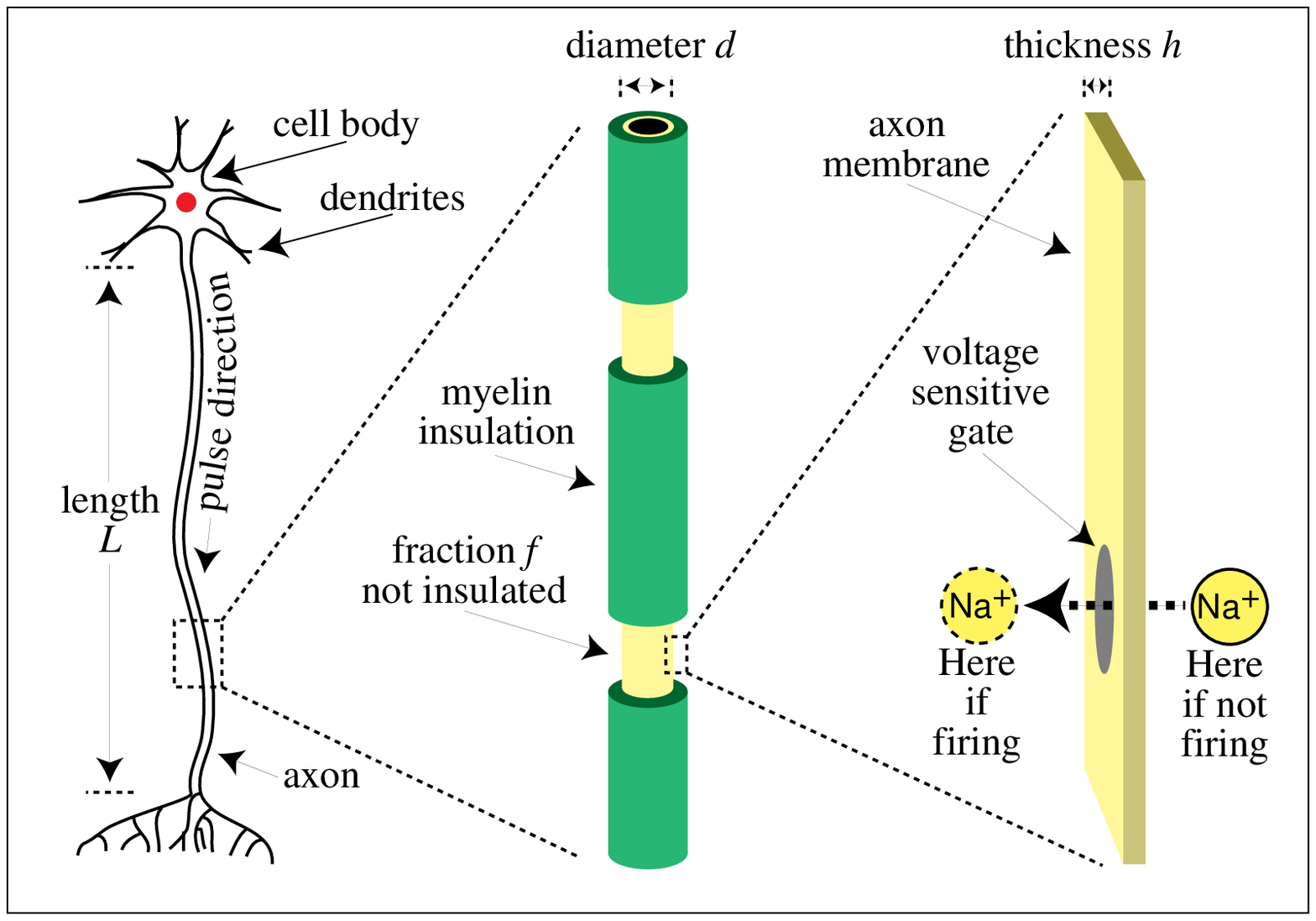}}
\bigskip
\mycaption{NeuronFig}{Schematic 
illustration of a neuron (left), 
a section of the myelinated axon (center) and 
and a piece of its axon membrane (right).
The axon is typically insulated (myelinated) with small 
bare patches every $0.5\,mm$ or so (so-called
Nodes of Ranvier) where the voltage-sensitive
sodium and potassium gates are concentrated
\cite{Morell,Hirano}.
If the neuron is in a superposition of firing and 
not firing, then $N\sim 10^6$ $\Naplus$ ions are in
a superposition of being inside and outside the cell
(right).
}
\end{figure}

Consider a small patch of the membrane, assumed to 
be roughly flat with uniform thickness $h$ as in
\fig{NeuronFig}. 
If there is an excess surface density $\pm\sigma$ of charge near
the inside/outside membrane surfaces, 
giving a voltage differential $U$ across the membrane,
then application of Gauss' law tells us that 
$\sigma=\epsilon_0 E$, where the electric field strength 
in the membrane is $E=U/h$ and $\epsilon_0$ is the vacuum permittivity.
Consider an axon of length $L$ and diameter $d$, with a fraction 
$f$ of its surface area bare (not insulated with myelin).
The total active surface area is thus $A=\pi d L f$, so
the total number of $\Naplus$ ions that migrate in during 
firing is 
\beq{Neq}
N = {A\sigma\over q} 
= {\pi d L f\epsilon_0(U_1-U_0)\over q h},
\eeq
where $q$ is the ionic charge ($q=q_e$,
the absolute value of the electron charge).
Taking values typical for central nervous system axons
\cite{Hirano,Ritchie},
%
%
$h=8\,nm$,
$d=10\,\um$, 
$L=10\,$cm,
$f=10^{-3}$,
$U_0=-0.07\,$V and
$U_1=+0.03\,$V
gives $N\approx 10^6$ ions, 
and reasonable variations in our parameters can change this 
number by a few orders of magnitude.

\subsection{Neuron decoherence mechanisms}

Above we saw that a quantum superposition of the neuron states
``resting'' and ``firing'' involves of order a million ions being 
in a spatial superposition of inside and outside the axon membrane, 
separated by a distance of order $h\sim 10\,\nm$.
In this subsection, we will compute the timescale on which 
decoherence destroys such a superposition.

In this analysis, the object is the neuron, and
the superposition will be destroyed by any interaction with 
other (environment) degrees of freedom that is sensitive to
where the ions are located. 
We will consider the following three sources of decoherence
for the ions:
\begin{enumerate}
\item Collisions with other ions
\item Collisions with water molecules
\item Coloumb interactions with more distant ions
\end{enumerate}
There are many more decoherence mechanisms 
\cite{ZehBook,JZ85,collapse}.
Exotic candidates such as 
quantum gravity \cite{Wormholes} and modified quantum mechanics
\cite{Pearle76,GRW} are generally much weaker \cite{collapse}.
A number of decoherence effects may be even stronger than 
those listed, \eg, interactions as the ions penetrate the
membrane --- the listed effects will turn out to be so strong 
that we can make our argument by simply using them
as lower limits on the actual decoherence rate.

Let $\rho$ denote the density matrix for the position 
$\r$ of a single $\Naplus$ ion.
As reviewed in the Appendix, all three of the listed processes
cause $\rho$ to evolve as
\beq{OffdiagSuppEq}
\rho(\x,\x',t_0+t) = \rho(\x,\x',t_0) f(\x,\x',t)
\eeq
for some function $f$ that is independent of the ion state $\rho$
and depends only on the interaction Hamiltonian $\Hint$.
This assumes that we can neglect the motion of the ion itself
on the decoherence timescale --- we will see that this condition
is met with a broad margin.

\subsubsection{Ion--ion collisions}

For scattering of environment particles (processes 1 and 2)
that have a typical de Broigle wavelength $\leff$, we have
\cite{collapse}
\beqa{fEq1}
f(\x,\x',t)
&=&e^{-\Lambda t\left(1-e^{-|\x'-\x|^2/2\leff^2}\right)}\nonumber\\
&\approx&
\cases{
e^{-|\x'-\x|^2 \Lambda t/2\leff^2}&for $|\x'-\x|\ll\leff$,\crr
e^{-\Lambda t}&for $|\x'-\x|\gg\leff$.
}
\eeqa
Here $\Lambda$ is the scattering rate, given by 
$\Lambda\equiv n\expec{\sigma v}$, where
$n$ is the density of scatterers, 
$\sigma$ is the scattering cross section 
and $v$ is the velocity. The product $\sigma v$ is
averaged over a the velocity distribution, which we
take to be a thermal (Boltzmann) distribution for 
corresponding to $T=37^\circ$C$\,\approx 310\,$K.
The gist of \eq{fEq1} is that a single collision 
decoheres the ion down to the de Broigle wavelength of
the scattering particle.
The information $I_{12}$ communicated during
the scattering is $I_{12}\sim\log_2(\Delta x/\lambda)$ bits,
where $\Delta x$ is the initial spread in the position of
our particle.

Since the typical de Broigle wavelength of a
$\Naplus$ ion 
(mass $m\approx 23 m_p$) or 
$\water$ molecule 
($m\approx 18 m_p$)
is 
\beq{deBroigleEq}
\lambda
= {2\pi\hbar\over\sqrt{3 m k T}}
\approx 0.03\,\nm
\eeq
at 310K, way smaller than the 
the membrane thickness $h\sim 10\,\nm$
over which we need to maintain quantum coherence, 
we are clearly in the $|\x'-\x|\gg\leff$ limit of 
\eq{fEq1}.
This means that the spatial superposition of an
ion decays exponentially 	
$\Lambda^{-1}$, of order its mean free time between collisions.
Since the superposition of the neuron states ``resting'' and ``firing''
involves $N$ such superposed ions, 
it thus gets destroyed on a timescale 
$\tau\equiv (N\Lambda)^{-1}$.

Let us now evaluate $\tau$.
Coulomb scattering between two ions of unit charge 
gives substantial deflection angles ($\theta\sim 1$)
with a cross section or order\footnote{
If the first ion starts at rest at $\r_1=(0,0,0)$ and 
the second is incident with $\r_2 = (vt,b,0)$,
then a very weak scattering with deflection angle
$\theta\ll 1$ will leave these trajectories
roughly unchanged, the radial force 
$F=g q^2/|\r_1-\r_2|^2$
merely causing 
a net transverse acceleration \cite{Jackson}
\beq{DeflectionEq}
\Delta v_y=\int_{-\infty}^{\infty}{\yh\cdot\F\over m} dt
= \int_{-\infty}^{\infty} {g q^2 b \,dt\over[b^2+(vt)^2]^{3/2}}
= {2 g q^2\over m v b}.
\eeq
The approximation breaks down as the
deflection angle $\theta\approx\Delta v_y/v$
approaches unity. This occurs for 
$b\sim g q^2/m v^2$, 
giving $\sigma=\pi b^2$ as in
\eq{CoulombSigmaEq}.
}
\beq{CoulombSigmaEq}
\sigma\sim\left({g q^2\over mv^2}\right)^2,
\eeq
where $v$ is the relative velocity and 
$g\equiv 1/4\pi\epsilon_0$ is the Coulomb constant.
In thermal equilibrium, the kinetic energy
$mv^2/2$ is of order 
$kT$,
so $v\sim \sqrt{kT/m}$.
For the ion density, let us write 
$n=\eta n_{\water}$, where
the density of water molecules $n_{\water}$ is about
$(1\,$g/$\cm^3)/(18m_p)\sim 10^{23}/\cm^3$
and $\eta$ is the relative concentration of
ions (positive and negative combined).
Typical ion concentrations during the resting state are
$[\Naplus]=$9.2 (120) mmol/l and
$[\Kplus]=$140 (2.5) mmol/l inside (outside) the axon membrane \cite{Katz},
corresponding to total $\Naplus+\Kplus$ concentrations 
of $\eta\approx 0.00027$ ($0.00022$) inside (outside).
To be conservative, we will simply use $\eta\approx 0.0002$ throughout.
Ion--ion collisions therefore destroy the superposition
on a timescale
\beq{IonTauEq}
\tau\sim {1\over N n \sigma v}
\sim{\sqrt{m (kT)^3}\over N g^2 q_e^4 n}
\sim 10^{-20}\,\seconds.
\eeq

\subsubsection{Ion--water collisions}

Since $\water$ molecules are electrically neutral, the cross-section
is dominated by their electric dipole moment 
$p\approx 1.85\,$Debye$\,\approx (0.0385\,\nm)\times q_e$.
We can model this dipole as two opposing unit charges 
separated by a distance $y\equiv p/q_e\ll b$, so 
summing the two corresponding contributions 
from \eq{DeflectionEq} gives a deflection 
angle 
\beq{DipoleDeflectionEq}
\theta \approx {2 g q_e p\over m v^2 b^2}.
\eeq
This gives a cross section
\beq{DipoleSigmaEq}
\sigma=\pi b^2 \sim{g q_e p\over m v^2}.
\eeq
for scattering with large $(\theta\sim 1)$ deflections.
Although $\sigma$ is smaller than for the 
case of ion--ion collisions, 
$n$ is larger because the concentration factor $\eta$
drops out, giving a final result
\beq{WaterTauEq}
\tau\sim {1\over N n \sigma v}
\sim{\sqrt{m k T}\over N g q_e p n}
\sim 10^{-20}\,\seconds
\eeq

\subsubsection{Interactions with distant ions}

As shown in the Appendix, long-range interaction 
with a distant (environment) particle gives
\beq{fEq2}
f(\r,\r',t) 
= \ph_2\left[\M(\r'-\r)t/\hbar\right],
\eeq
up to a phase factor that is irrelevant for 
decoherence.
Here $\ph_2$ is the Fourier transform of 
$p_2(\r)\equiv\rho_2(\r,\r)$, the probability
distribution for the location of the environment particle. 
$M$ is the $3\times 3$ Hessian matrix of second derivatives
of the interaction potential of the two particles at 
their mean separation. 
A slightly less general formula was derived in the 
seminal paper \cite{JZ85}.
For roughly thermal states, $\rho_2$ (and thus $p$) is likely to be
well approximated by a Gaussian \cite{ZHP,gaussians}. This gives
\beq{fEq3}
f(\r,\r',t) 
= e^{-{1\over 2}(\r'-\r)^t\M^t\Sig\M(\r'-\r)t^2/\hbar^2},
\eeq
where
$\Sig=\expec{\r_2\r_2^t}-\expec{\r_2}\expec{\r_2^t}$
is the covariance matrix of the location of the environment particle.
Decoherence is destroyed when the exponent becomes of order unity, 
\ie, on a timescale
\beq{JZtauEq}
\tau\equiv\left[(\r'-\r)^t\M^t\Sig\M(\r'-\r)\right]^{-1/2}\hbar.
\eeq
Assuming a Coulomb potential $V=g q^2/|\r_2-\r_1|$
gives $\M=(3\ah\ah^t-\I)g q^2/a^3$ where 
$\a\equiv\r_2-\r_1=a\ah$, $|\ah|=1$.
For thermal states, we have the isotropic case $\Sig=(\Delta x)^2\I$,
so \eq{JZtauEq} reduces to
\beq{JZtauEq2}
\tau={\hbar a^3\over g q^2|\r'-\r|\Delta x}\left(1+3\cos^2\theta\right)^{-1/2},
\eeq
where $\cos\theta\equiv\ah\cdot(\r'-\r)/|\r'-\r|$.
To be conservative, we take $\Delta x$ to be as small as the uncertainty
principle allows.
With the thermal constraint $(\Delta p)^2/m\simlt kT$ on the momentum
uncertainty, this gives
\beq{DxEq}
\Delta x = {\hbar\over 2\Delta p}\sim {\hbar\over\sqrt{m k T}}.
\eeq
Substituting this into \eq{JZtauEq2} and dividing by the number
of ions $N$, we obtain the decoherence
timescale
\beq{JZtauEq3}
\tau\sim{a^3\sqrt{m k T}\over N g q^2 |\r'-\r|}.
\eeq
caused by a single environment ion a distance $a$ away.
Each such ion will produce its own suppression factor
$f$, so we need to sum the exponent in \eq{fEq3} over all ions.
Since the tidal force $\M\propto a^{-3}$ causes the exponent to 
drop as $a^{-6}$,
this sum will generally be dominated by the very closest ion,
which will typically be a distance $a\sim n^{-1/3}$ away.
We are interested in decoherence for separations 
$|\r'-\r|=h$, the membrane thickness, which gives 
\beq{JZtauEq4}
\tau\sim{\sqrt{m k T}\over N g q_e^2 n h}\sim 10^{-19}\,\seconds.
\eeq 
The relation between these different estimates is discussed in 
more detail in the Appendix.

\subsection{Microtubules}

Microtubules are a major component of the cytoskeleton, 
the ``scaffolding'' that helps cells maintain their shapes.
They are hollow cylinders of diameter $D=24\,nm$ made up of 13 
filaments 
that are strung together out of 
proteins known as tubulin dimers. These dimers can make transitions
between two states known as $\alpha$ and $\beta$, corresponding 
to different electric dipole moments along the axis of the tube.
It has been argued that microtubules may have additional functions
as well, serving as a means of energy and information 
transfer \cite{Hameroff}.
A model has been presented 
whereby the dipole-dipole interactions
between nearby dimers can lead to long-range 
polarization and kink-like excitations that may travel down the
microtubules at speeds exceeding 1 m/s \cite{Sataric}.

Penrose has gone further and suggested that the dynamics 
of such excitations 
can make a microtubule act like a quantum computer, 
and that microtubules are the site of of human consciousness 
\cite{Penrose97}.
This idea has been further elaborated
\cite{Nano95,Mavro95a,Mavro95b,Mavro95c} 
employing methods from string theory,
with the conclusion that quantum superpositions of 
coherent excitations can persist for as long as a second before 
being destroyed by decoherence. See also \cite{Rosu96a,Rosu96b}.
This was hailed 
as a success for the model, the interpretation being 
that the quantum gravity effect on microtubules was identified with
the human though process on this same timescale.

This decoherence rate $\tau\sim 1\,\seconds$ was computed assuming 
that quantum gravity is the main decoherence source.
Since this quantum gravity model is 
somewhat controversial \cite{Hawking97} and its effect
has been found to be 
more than 20 orders of magnitude weaker than other decoherence
sources in some cases \cite{collapse}, it 
seems prudent to evaluate other decoherence sources for the
microtubule case as well, to see whether they are in fact dominant.
We will now do so.

Using coordinates where the $x$-axis is along the tube axis, 
the above-mentioned models all focus
on the time-evolution of $p(x)$, the average x-component of the
electric dipole moment of the tubulin dimers at each $x$.
In terms of this polarization function $p(x)$, the net charge 
per unit length of tube is $-p'(x)$. 
The propagating kink-like excitations \cite{Sataric}
are of the form
\beq{KinkEq}
p(x) = 
\cases{
+p_0 &for $x\ll x_0$,\crr
-p_0 &for $x\gg x_0$,
}
\eeq
where the 
kink location $x_0$ propagates with constant speed and has
a width of order a few tubulin dimers.
The polarization strength $p_0$ is such that the 
total charge around the kink is
$Q = -\int p'(x)dx = 2 p_0\sim 940 q_e$,
due to the presence of 18 Ca$^{2+}$ ions on each of the 
13 filaments contributing to $p_0$ \cite{Sataric}.

Suppose that such a kink is in two different places in superposition,
separated by some distance $|\r'-\r|$. 
How rapidly will the superposition be destroyed by decoherence?
To be conservative, we will ignore collisions between polarized
tubulin dimers and nearby water molecules, since it has been 
argued that these may be in some sense ordered and part of 
the quantum system \cite{Mavro95c}
-- although this argument is difficult to maintain for the 
water outside the microtubule, which permeates the entire 
cell volume.
Let us instead apply \eq{JZtauEq3}, with 
$N=Q/q_e\sim 10^3$.
The distance to the nearest ion will generally be less than
$a = R + n^{-1/3}\sim 26\,\nm$,
where the tubulin diameter $D=24\,\nm$
dominates over the inter-ion separation 
$n^{-1/3}\sim 2\,\nm$ in the fluid surrounding the
microtubule.
Superpositions spanning many tubuline dimers
($|\r'-\r|\gg D$) therefore decohere on a timescale
\beq{JZtauEq5}
\tau\sim{D^2\sqrt{m k T}\over N g q_e^2}\sim 10^{-13}\,\seconds.
\eeq
due to the nearest ion alone. This is quite a conservative
estimate, since the other
$n D^3\sim 10^3$ ions that are merely
a small fraction further away will also contribute to the decoherence rate,
but it is nonetheless 6-7 orders of magnitude shorter than the
estimates of Mavromatos \& Nanopoulos \cite{Mavro98a,Mavro98b,Mavro99}.
We will comment on screening effects below.

\subsubsection{Decoherence summary}

Our decoherence rates are summarized in Table 1.
How accurate are they likely to be?

In the calculations above, we  
generally tried to be conservative, erring on the side of 
underestimating the decoherence rate. 
For instance, we neglected that $N$ potassium 
ions also end up 
in superposition once the neuron firing is quenched,
we neglected the contribution of other abundant ions such as 
Cl$^-$ to $\eta$, and
and we ignored collisions with water molecules in the microtubule case.

Since we were only interested in order-of-magnitude
estimates, we made a number of crude approximations, 
\eg, for the cross sections.
We neglected screening effects 
because the decoherence rates were dominated by the particles
closest to the system, \ie, the very same particles 
that are responsible for screening the charge from more distant ones.

\def\filler{\hbox{$\quad\quad$}}
\begin{center}
{\footnotesize
{\bf Table 1.} Decoherence timescales.
\noindent
\begin{tabular}{llc}
\noalign{\vskip 4pt}
\hline
\hline
Object  	&Environment		&$\tdec$\\
\hline	
\noalign{\vskip 2pt}
Neuron		&Colliding ion		&$10^{-20}$s\\
Neuron		&Colliding $\water$\filler&$10^{-20}$s\\
Neuron		&Nearby ion		&$10^{-19}$s\\
Microtubule\filler&Distant ion		&$10^{-13}$s\\
\hline
\hline
\end{tabular}
}
\end{center}

\section{Discussion}
\label{DiscussionSec}

\subsection{The classical nature of brain processes}

The calculations above enable us to address the question of whether 
cognitive processes in the brain constitute
a classical or quantum system in the sense 
of \fig{TimescaleFig}.
If we take the characteristic dynamical timescale for such processes
to be $\tdyn\sim 10^{-2}\,\seconds-10^{0}\,\seconds$ (the apparent timescale of
\eg, speech, thought and motor response), 
then a comparison of $\tdyn$ with $\tdec$ from Table 1 
shows that processes associated with either 
conventional neuron firing or with 
polarization excitations in microtubules
fall squarely in the classical category, 
by a margin exceeding ten orders of magnitude.
Neuron firing itself is also highly classical, since it occurs
on a timescale $\tdyn\sim 10^{-3}-10^{-4}\,\seconds$
\cite{Ritchie}.
Even a kink-like microtubule excitation is classical 
by many orders of magnitude, since it traverses a short
tubule on a timescale $\tdyn\sim 5\times 10^{-7}\,\seconds$
\cite{Sataric}.

What about other mechanisms?
It is worth noting that if (as is commonly believed)
different neuron firing patterns correspond in some
way to different conscious perceptions, then consciousness itself
cannot be of a quantum nature even if there is a yet undiscovered
physical process in the brain with a very long decoherence
time. As mentioned above, suggestions for such candidates have involved,
\eg, superconductivity \cite{Walker},
superfluidity \cite{Domash},
electromagnetic fields \cite{Stapp93},
Bose condensation \cite{Marshall,Zohar},
superflourescence \cite{Rosu} and 
other mechanisms \cite{Umezawa,Vitiello95}.
The reason is that as soon as such a quantum subsystem 
communicates with the constantly decohering neurons to create 
conscious experience, everything decoheres.

How extreme variations in the decoherence rates can we obtain by
changing our model assumptions?
Although the rates can be altered by a few
of orders of magnitudes by pushing parameters such as the
neuron dimensions, the myelination fraction or the microtubule kink charge
to the limits of plausibility, it is clearly impossible to 
change the basic conclusion that
$\tdec\ll 10^{-3}\,\seconds$, \ie, that we are dealing with
a classical system in the sense of \fig{TimescaleFig}.
Even the tiniest neuron imaginable, with only a single 
ion ($N=1$) traversing the cell wall during firing, 
would have $\tdec\sim 10^{-14}\,\seconds$.
Likewise, reducing the effective microtubule kink charge to a small
fraction of $q_e$ would not help.

How are we to understand the above-mentioned claims that brain subsystems 
can be sufficiently isolated to exhibit macroquantum behavior?
It appears that the subtle distinction between 
dissipation and decoherence timescales has not always been appreciated.

\subsection{Implications for the subject-object-environment decomposition}

Let us now discuss the subsystem decomposition of \fig{TrinityFig}
in more detail in light of our results.
As the figure indicates, the virtue of this decomposition
into subject, object and environment is that 
the subsystem Hamiltonians 
$\Hsubj$, $\Hobj$, $\Henv$ and the interaction Hamiltonians
$\Hso$, $\Hoe$, $\Hse$ can
cause qualitatively very different effects.
Let us now briefly discuss each of them in turn.

Most of these processes are schematically
illustrated in \fig{ChessFig} and \fig{ChessFig2}, where for 
purposes of illustration,
we have shown the extremely simple case where both the subject and object
have only a single degree of freedom that can take on only a few
distinct values (3 for the subject, 2 for the object).
For definiteness, 
we denote the three subject states 
$\noobs$, $\upobs$ and $\downobs$,
and interpret them as the observer
feeling neutral, happy and sad, respectively.
We denote the two object states $\up$ and $\down$,
and interpret them as 
the spin component (``up'' or ``down'') 
in the $z$-direction of a spin-1/2 system,
say a silver atom.
The joint system consisting of subject and object therefore has
only $2\times 3=6$ basis states:
$\noup$, $\nodown$,
$\upup$, $\updown$,
$\downup$, $\downdown$.
In Figures~\arabic{ChessFig} and~\arabic{ChessFig2}, we have 
therefore plotted $\rho$
as a $6\times 6$ matrix consisting of nine two-by-two blocks.

\begin{figure}[tb] 
\centerline{\epsfxsize=3.5in\epsffile{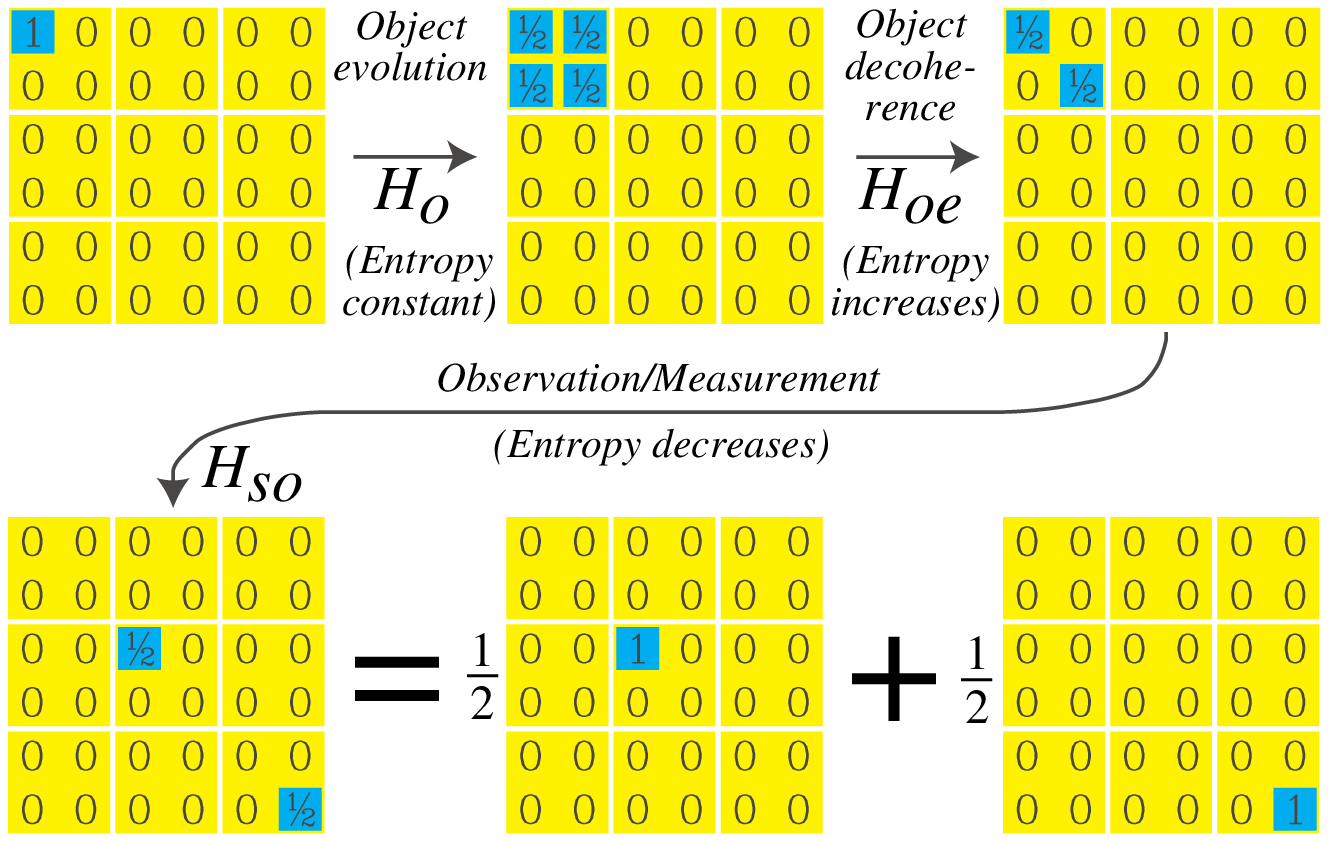}}
\smallskip
\mycaption{ChessFig}{Time evolution 
of the $6\times6$ density matrix
for the basis states 
$\noup$, $\nodown$,
$\upup$, $\updown$,
$\downup$, $\downdown$ 
as the object evolves in isolation, then decoheres,
then gets observed by the subject. The final result
is a statistical mixture of the states
$\upup$ and $\downdown$, simple zero-entropy states
like the one we started with.
}
\end{figure}

\subsubsection{Effect of $\Hobj$: constant entropy}

If the object were to evolve during a time interval $t$
without interacting with
the subject or the environment ($\Hso=\Hoe=0$), 
then according to \eq{HeisenbergEq}
its reduced density matrix $\robj$ would evolve
into $U\robj U^\dagger$ with the same entropy,
since the time-evolution operator 
$U\equiv e^{-i\Hobj t}$ is unitary.

Suppose the subject stays in the state $\noobs$ and the object
starts out in the pure state $\up$.
Let the object Hamiltonian
$\Hobj$ correspond to a magnetic field in the $y$-direction
causing the spin to precess to the $x$-direction, \ie, to the state
$(\up+\down)/\sqrt{2}$.
The object density matrix $\robj$ then evolves into
\beqa{ConstantEntropyEq}
\robj 
&=&U\up\upbra U^\dagger = {1\over 2}(\up+\down)(\upbra+\downbra)\nonumber\\
&=&{1\over 2}(\up\upbra + \up\downbra + \down\upbra + \down\downbra),
\eeqa
corresponding to the four entries of $1/2$ in 
the second matrix of \fig{ChessFig}.

This is quite typical of pure quantum time evolution:
a basis state eventually evolves into a superposition
of basis states, 
and the quantum nature of this superposition
is manifested by off-diagonal elements in $\robj$. 
Another familiar example of this is 
the familiar spreading out of the wave packet of 
a free particle.

\subsubsection{Effect of $\Hoe$: increasing entropy}

This was the effect of $\Hobj$ alone. 
In contrast, $\Hoe$ will generally cause decoherence and increase 
the entropy of the object. 
As discussed in detail in \sec{CalculationSec} and the Appendix,
it entangles it with the environment, 
which suppresses the off-diagonal elements of
the reduced density matrix of the object as illustrated
in \fig{ChessFig}. If $\Hoe$ couples to the $z$-component of the
spin, this 
destroys the terms $\up\downbra$ and $\down\upbra$.
Complete decoherence therefore converts the final state of 
\eq{ConstantEntropyEq} into 
\beq{IncreasingEntropyEq}
\robj = {1\over 2}(\up\upbra + \down\downbra),
\eeq
corresponding to the two entries of $1/2$ in the
third matrix of \fig{ChessFig}.

\subsubsection{Effect of $\Hso$: decreasing entropy}

Whereas $\Hoe$ typically causes the apparent 
entropy of the object to increase,
$\Hso$ typically causes it to decrease.
\Fig{ChessFig} illustrates the case of an ideal measurement, where
the subject starts out in the state $\noobs$
and $\Hso$ is of such a form that 
gets perfectly correlated with the object.
In the language of \sec{SubsystemSec},
an ideal measurement is a type of 
communication where the mutual information 
$I_{12}$ between the subject and object systems
is increased to its maximum possible value.
Suppose that the measurement is caused by $\Hso$ becoming large during 
a time interval so brief that we can neglect the effects of 
$\Hsubj$ and $\Hobj$. The joint subject$+$object density matrix
$\rso$ then evolves as $\rso\mapsto U\rso U^\dagger$,
where $U\equiv\exp\left[-i\int\Hso dt\right]$.
If observing $\up$ makes the subject happy and 
$\down$ makes the subject sad, then we have
$U\noup=\upup$ and $U\nodown=\downdown$.
The state given by \eq{IncreasingEntropyEq}
would therefore evolve into
\beqa{DecreasingEntropyEq}
\robj
&=&{1\over 2}U(\noobs\noobsbra)\tensormult(\up\upbra + \down\downbra)U^\dagger\\ 
&=&{1\over 2}(U\noup\noupbra U^\dagger + U\nodown\nodownbra U^\dagger\\ 
&=&{1\over 2}(\upup\upupbra + \downdown\downdownbra),
\eeqa
as illustrated in \fig{ChessFig}.
This final state contains a mixture of 
two subjects, corresponding to definite but opposite 
knowledge of the object state.
According to both of them, the entropy of the object has
decreased from one bit to zero bits.

In general, we see that the object decreases its entropy
when it exchanges information with the subject and increases 
when it exchanges information with the environment.\footnote{If 
$n$ bits of information are exchanged with the environment, then
\eq{InfoDefEq} shows that
the object entropy will increase by this same amount
if the environment is in thermal equilibrium (with maximal entropy) 
throughout. If we were to know the state of the environment
initially (by our definition of environment, we do not), then 
both the object and environment entropy will typically increase
by $n/2$ bits.}
Loosely speaking, the entropy of an object 
decreases while you look at it and 
increases while you don't\footnote{
Here and throughout, we are assuming that 
the total system, which is by definition isolated,
evolves according to the 
Schr\"odinger \eq{HeisenbergEq}.
Although modifications of the Schr\"odinger
equation have been suggested by some authors,
either in a mathematically explicit form
as in \cite{Pearle76,GRW} or verbally as a
so-called reduction postulate, there
is so far no experimental evidence 
suggesting that modifications are necessary.
The original motivations for such modifications were
\begin{enumerate}
\item to be able to interpret the diagonal elements of 
the density matrix as probabilities and
\item to suppress off-diagonal elements of the density
matrix.
\end{enumerate}
The subsequent discovery by Everett \cite{Everett} that
the probability interpretation automatically 
appears to hold for almost all observers in the final 
superposition solved problem 1, and is discussed in more detail in, \eg,
\cite{Zeh81,Lockwood,Deutsch,Page95,Donald97,Donald99,Vaidman,Sakaguchi,everett,TOE}.
The still more recent discovery of decoherence 
\cite{Zeh70,Zurek81,Zurek82}
solved problem 2,
as well as explaining so-called superselection rules
for the first time (why for instance the position basis has
a special status) \cite{ZehBook}.
}.

\begin{figure}[tb] 
\centerline{\epsfxsize=3.5in\epsffile{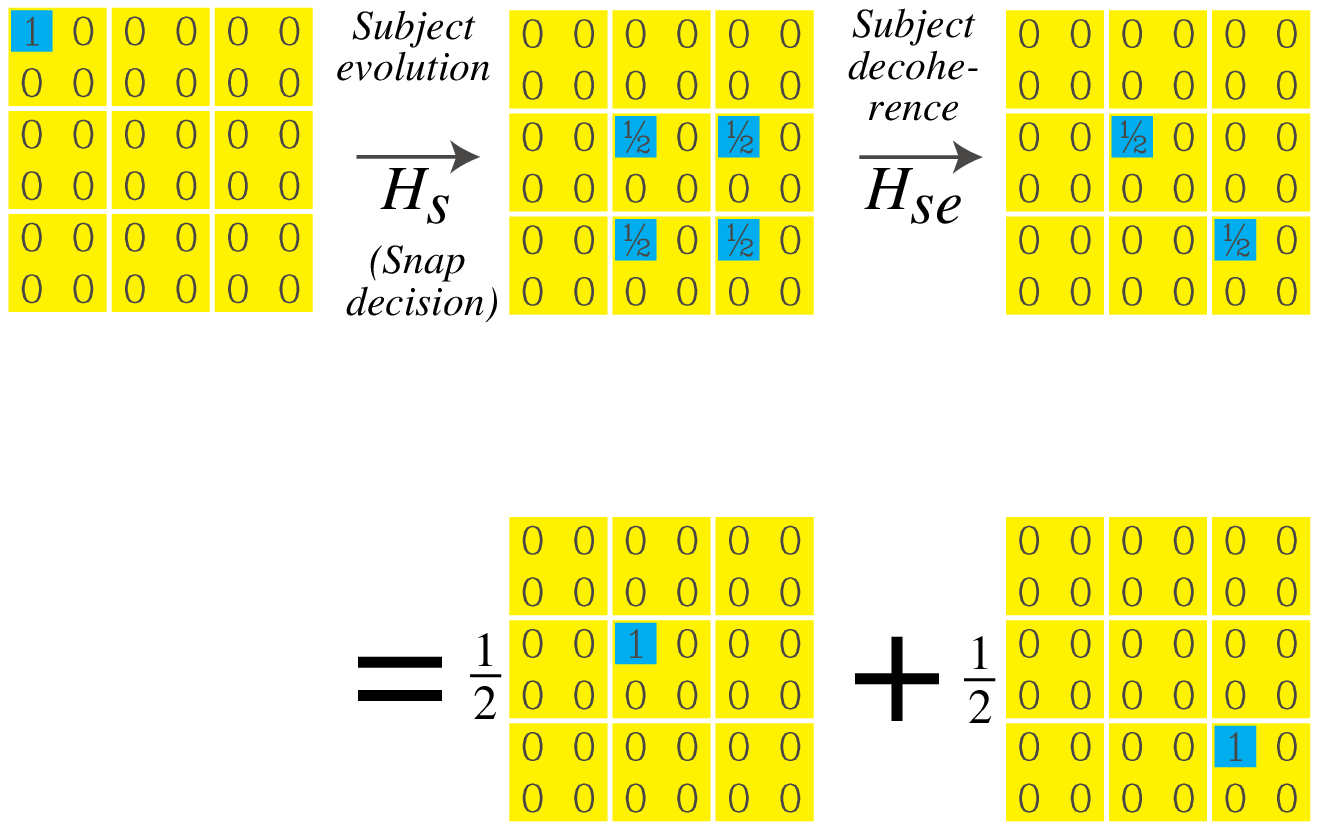}}
\smallskip
\mycaption{ChessFig2}{Time evolution 
of the same $6\times6$ density matrix
as in \fig{ChessFig}
when the subject evolves in isolation, then decoheres.
The object remains in the state $\up$ the whole time.
The final result
is a statistical mixture of the two states
$\upup$ and $\downup$.
}
\end{figure}

\subsubsection{Effect if $\Hsubj$: the thought process}

So far, we have focused on the object and discussed effects of
its internal dynamics ($\Hobj)$ and 
its interactions with the environment ($\Hoe$) and 
subject ($\Hso$). Let us now turn to the subject and
consider the role played by its internal dynamics
($\Hsubj)$ and interactions with the environment ($\Hse$).
In his seminal 1993 book, Stapp \cite{StappBook} 
presents an argument about brain dynamics that can be summarized
as follows.
\begin{enumerate}
\item 
Since the brain contains $\sim 10^{11}$ 
synapses connected together 
by neurons in a highly nonlinear fashion, there must be
a huge number of metastable reverberating patters of pulses into 
which the brain can evolve.
\item Neural network simulations have indicated that the metastable state into which
a brain does in fact evolves depends sensitively on the initial conditions in
small numbers of synapses.
\item The latter depends on the locations of a small number of calcium atoms,
which might be expected to be in quantum superpositions.
\item Therefore, one would expect the brain to evolve into 
a quantum superposition of many such metastable configurations.
\item 
Moreover, the fatigue characteristics of the synaptic junctions
will cause any given metastable state to become, after a short time, 
unstable: the subject will then be forced to search for a new metastable 
configuration, and will therefore continue to evolve into a superposition of
increasingly disparate states.
\end{enumerate}
If different states (perceptions) of the subject correspond to 
different metastable states of neuron firing patterns,
a definite perception would eventually evolve
into a superposition of several subjectively distinguishable perceptions.

We will follow Stapp in making this assumption about $\Hsubj$.
For illustrative purposes, let us assume that this can happen even 
at the level of a single thought or snap decision where the outcome
feels unpredictable to us.
Consider the following experiment: 
the subject starts out with a blank face 
and counts silently to three, 
then makes a snap decision on whether to smile or frown.
The time-evolution operator $U\equiv\exp\left[-i\int\Hsubj dt\right]$
will then have the property
that $U\noobs = (\upobs+\downobs)/\sqrt{2}$,
so the subject density matrix $\rsubj$ will evolve into 
\beqa{SnapDecisionEq}
\rsubj 
&=&U\noobs\noobsbra U^\dagger 
= {1\over 2}(\upobs+\downobs)(\upobsbra+\downobsbra)\nonumber\\
&=&{1\over 2}(\upobs\upobsbra + \upobs\downobsbra + \downobs\upobsbra + \downobs\downobsbra),
\eeqa
corresponding to the four entries of $1/2$ in the second matrix in 
\fig{ChessFig2}.

\subsubsection{Effect of $\Hse$: subject decoherence}

Just as $\Hoe$ can decohere the object, 
$\Hse$ can decohere the subject. 
The difference is that whereas the object can be either
a quantum system (with small $\Hoe$) or a classical
system (with large $\Hoe$), a human subject
{\it always} has a large interaction with the environment.
As we showed in \sec{CalculationSec}, 
$\tdec\ll\tdyn$ for the subject, \ie, the
effect of $\Hse$ is faster than that 
of $\Hsubj$ by many orders of magnitude.
This means that we should strictly speaking not think of 
macrosuperpositions such as \eq{SnapDecisionEq}
as first forming and then decohering as in \fig{ChessFig2}
--- rather, subject decoherence is so fast that
such superpositions decohere already 
during their process of formation.
Therefore we are never even close to being able to 
perceive superpositions of different perceptions.
Reducing object decoherence (from $\Hoe$) 
during measurement would make no difference,
since decoherence would take place in the brain
long before the transmission of the appropriate 
sensory input through sensory nerves had 
been completed.

\subsection{$H_e$ and $H_{soe}$}

The environment is of course the most complicated system, since
it contains the vast majority of the degrees of freedom 
in the total system. It is therefore very fortunate that we 
can so often ignore it, considering only those limited
aspects of it that affect the subject and object.

For the most general $H$, there can 
also be an ugly irreducible residual term 
$H_{soe}\equiv H-\Hsubj-\Hobj-\Henv-\Hso-\Hoe-\Hse$.

\subsection{Implications for modeling cognitive processes}

For the neural network community, 
the implication of our result is 
``business as usual'', \ie, there is no
need to worry about the fact that current simulations
do not incorporate effects of quantum coherence.
The only remnant from quantum mechanics is 
the apparent randomness that we subjectively 
perceive every time the subject system 
evolves into a superposition as in \eq{SnapDecisionEq},
but this can be simply modeled by including 
a random number generator in the simulation.
In other words, the recipe used to prescribe when 
a given neuron should fire and how synaptic
coupling strengths should be updated
may have to involve some classical randomness
to correctly mimic the behavior of the brain.

\subsubsection{Hyper-classicality}

If a subject system is to be a good model of us,
$\Hso$ and $\Hse$ need to meet certain criteria:
decoherence and communication are necessary, 
but fluctuation and dissipation must be kept low enough that the
subject does not lose its autonomy completely.

In our study of neural processes, we concluded that the 
subject is not a quantum system, since $\tdec\ll\tdyn$.
However, since the dissipation time $\tdiss$ for neuron firing
is of the same order as its dynamical timescale, we see
that in the sense of \fig{TimescaleFig}, the subject is not a simple 
classical system either.
It is therefore somewhat misleading to think of it as 
simply some classical degrees of freedom evolving fairly undisturbed
(only interacting enough to stay decohered and occasionally 
communicate with the outside world).
Rather, the semi-autonomous degrees of freedom that 
constitute the subject are to be found at a higher level
of complexity, perhaps as metastable 
global patters of neuron firing.

These degrees of freedom might be termed ``hyper-classical'':
although there is nothing quantum-mechanical about
their equations of motion (except that they can be stochastic),
they may bear little resemblance with the underlying 
classical equations from which they were derived.
Energy conservation and other familiar concepts from
Hamiltonian dynamics will be irrelevant for these more
abstract equations, since neurons are energy 
pumped and highly dissipative.
Other examples of such hyper-classical systems
include the time-evolution of the memory contents of 
a regular (highly dissipative) digital computer
as well as the motion on the screen of objects in
a computer game.

\subsubsection{Nature of the subject system}

In this paper, we have tacitly assumed that consciousness is 
synonymous with certain brain processes.
This is what Lockwood terms the ``identity theory'' \cite{Lockwood}.
It dates back to Hobbes ($\sim$1660) and has been
espoused by, {\eg}, Russell, Feigl, Smart, Armstrong,
Churchland and Lockwood himself.
Let us briefly explore the more specific assumption that
the subject degrees of freedom {\it are} our perceptions.
In this picture, some of the subject degrees of freedom 
would have to constitute a ``world model'', with
the interaction $\Hso$ such that the resulting communication
keeps these degrees of freedom highly correlated
with selected properties of the outside world 
(object $+$ environment).
Some such properties, \ie,
\begin{itemize}
\item the intensity of the electromagnetic 
on the retina,  
averaged through three narrow-band filters (color vision)
and one broad-band filter (black-and-white vision),
\item the spectrum of air pressure fluctuations in the ears 
(sound),
\item the chemical composition of gas in the nose (smell) and 
solutions in the mouth (taste),
\item heat and pressure at a variety of skin locations,
\item locations of body parts,
\end{itemize}
are tracked rather continuously, with the corresponding 
mutual information $I_{12}$ between subject and surroundings
remaining fairly constant. 
Persisting correlations with properties of the past state
of the surroundings (memories) further 
contribute to the mutual information $I_{12}$. 
Much of $I_{12}$ is due to correlations
with quite subtle aspects of the surroundings, \eg, 
the contents of books.
The total mutual information $I_{12}$ between a
person and the external world is fairly low at birth, 
gradually grows through learning, and falls when we forget.
In contrast, most innate objects have a very small 
mutual information with the rest of the world, 
books and diskettes being notable exceptions.

The extremely limited selection of 
properties that the subject correlates with has
presumably been determined by evolutionary utility,
since it is known to differ between species:
birds perceive four primary colors but cats only one,
bees perceive light polarization, {\it etc}.
In this picture, we should therefore not consider these
particular (``classical'') aspects of our
surroundings to be more fundamental than the
vast majority that the subject system is
uncorrelated with.
Morover, our perception of {\eg} space is 
as subjective as our perception of color,
just as suggested by \eg \cite{Cairns-Smith}.

\subsubsection{The binding problem}

One of the motivations for models with quantum coherence
in the brain was the so-called binding problem.
In the words of James \cite{JamesBook,James},
``the only realities are the separate molecules, or at most cells.
Their aggregation into a `brain' is a fiction of popular speech''.
James' concern, shared by many after him, was
that consciousness did not seem to be spatially localized
to any one small part of the brain, yet subjectively
feels like a coherent entity.
Because of this, Stapp \cite{StappBook} and many others have 
appealed to quantum coherence, arguing that
this could make consciousness a holistic effect
involving the brain as a whole.

However, non-local degrees of freedom 
can be important even in classical physics,
For instance, oscillations in a guitar string are 
local in Fourier space, not in real space,
so in this case the ``binding problem'' can be solved
by a simple change of variables.
As Eddington
remarked \cite{Eddington}, when observing the ocean 
we perceive the moving waves as objects in their own right
because they display a certain permanence, 
even though the water itself is only bobbing up and down. 
Similarly, thoughts are presumably highly non-local excitation patterns
in the neural network of our brain, except of a non-linear and much
more complex nature.
In short, this author feels that there is no binding problem.

\subsubsection{Outlook}

In summary, our decoherence calculations have indicated that
there is nothing fundamentally 
quantum-mechanical about cognitive processes in the brain,
supporting the Hepp's conjecture \cite{Hepp99}.
Specifically, the {\it computations} in the brain appear
to be of a classical rather than quantum nature,
and the argument by Lisewski \cite{Lisewski}
that quantum corrections may
be needed for accurate modeling of some details,
\eg, non-Markovian noise in neurons, does of course 
not change this conclusion.
This means that although the current state-of-the-art 
in neural network hardware 
is clearly still very far from being able to 
model and understand cognitive processes as 
complex as those in the brain, there are
no quantum mechanical reasons to doubt that  
this research is on the right track.

\bigskip
{\bf Acknowledgements:}
The author wishes to thank the 
organizers of the Spaatind-98 and Gausdal-99 winter schools,
where much of this work was done,
and Mark Alford, 
Philippe Blanchard, 
Carlton Caves, 
Angelica de Oliveira-Costa,
Matthew Donald,
Andrei Gruzinov, 
Piet Hut, 
Nick Mavromatos, 
Henry Stapp, 
Hans-Dieter Zeh and 
Woitek Zurek for 
stimulating discussions and helpful comments.
Support for this work was provided by the Sloan Foundation and by
NASA though grant NAG5-6034 and 
Hubble Fellowship HF-01084.01-96A from STScI, operated by AURA, 
{\frenchspacing Inc.} 
under NASA contract NAS5-26555.

\appendix

\section*{Decoherence formulas}
\label{Appendix}

The quantitative effect of decoherence from both 
short range interactions (scattering) and long-range
interactions was first derived in a seminal 
paper by Joos \& Zeh \cite{JZ85}.
Since our application involved scattering between particles
of comparable mass, we used a generalized version 
of these results that included the effect of recoil \cite{collapse}.
In this Appendix, we derive a slightly generalized formula for
long-range interactions, and briefly comment on the relation 
between these short-range and long-range limiting cases. 

\subsection{Decoherence due to tidal forces}

Even if the dissipation and fluctuation caused by $\Hint$ is 
dynamically unimportant, 
$H_1$ and $H_2$ can be neglected in \eq{HintEq} when calculating
the decoherence effect in the many cases where the interaction Hamiltonian
decoheres the object on a timescale far below the dynamical time.
In this approximation,
we consider two particles with an interaction
$H = \Hint = V(\r_2-\r_1)$ for some potential $V$. 
According to \eq{HeisenbergEq}, the two-particle density matrix 
$\rho$ therefore evolves as 
\beqa{JZevolEq}
&&\rho(\r_1,\r_1',\r_2,\r_2',t_0+t)\nonumber\\
&=& \rho(\r_1,\r_1',\r_2,\r_2',t)
e^{-i[V(\r_2-\r_1)-V(\r'_2-\r'_1)]/\hbar}.
\eeqa
Following \cite{JZ85}, we
assume that the two particles are fairly localized near their 
initial average positions
\beq{r0defEq}
\r_i^0\equiv \expec{\r_i}_0 = \tr[\r_i\rho_i(t_0)],
\eeq
$i=1,2$,
and approximate the potential by its second order
Taylor expansion
\beqa{TaylorEq}
V(\r_2-\r_1)&\approx&V(\a) - \F\cdot(\x_2-\x_1)\nonumber\\
&+& {1\over 2}(\x_2-\x_1)^t\M (\x_2-\x_1).
\eeqa
Here $\F\equiv=-\nabla V(\a)$ is the average force, $\M$ is the 
Hessian matrix
$\M_{ij}\equiv\partial_i\partial_j V(\a)$ and
$\a\equiv\r_2^0-\r_1^0$.
We have introduced relative coordinates
$\x_i\equiv \r_i-\r_i^0$.
Assuming that the two particles are independent initially
as in \cite{JZ85}, \ie, that $\rho(t_0)$ takes the separable form 
$\rho(\x_1,\x_1',\x_2,\x_2',t_0) = 
\rho_1(\x_1,\x_1',t_0)\rho_2(\x_2,\x_2',t_0)$,
this gives
\beqa{ReducedJZeq}
&&\rho_1(\x_1,\x_1',t_0+t) = \tr_2 \rho(t_0+t) =\nonumber\\ 
&&\int\rho(\x_1,\x_1',\x,\x,t_0+t)d^3x = 
\rho_1(\x_1,\x_1',t_0) f(\x_1,\x_1',t),
\eeqa
where 
\beqa{JZfEq}
&&f(\x_1,\x_1',t)\approx\nonumber\\
&&e^{i\phi(\x_1,\x_1',t)} 
\int \rho_2(\x_2,\x_2',t_0) e^{-it(\x'_1-\x_1)^t\M\x_2/\hbar} d^3x_2 =\nonumber\\
&&e^{i\phi(\x_1,\x_1',t)}\ph_2[\M(\x_1'-\x_1)t/\hbar].
\eeqa
Here the phase factor 
\beq{phiEq}
e^{i\phi(\x,\x',t)}\equiv 
e^{{i\over\hbar}\left[\F\cdot(\x'-\x)
+{1\over 2}{\x'}^t\M\x'
-{1\over 2}{\x}^t\M\x\right]}
\eeq
is of no importance for decoherence, since it does not suppress
the magnitude 
$|\rho_1(\x_1,\x_1',t)|$ of the off-diagonal elements -- it merely 
causes momentum transfer related to fluctuation and dissipation.
It is the other term that causes decoherence.
$\ph_2$ is the Fourier transform of 
$p_2(\x)\equiv\rho_2(\x,\x,t_0)$, the probability distribution
for the location of the environment particle.

\subsection{Properties of the effect}

Let us briefly discuss some qualitative features of 
\eq{JZfEq}.
Since $\ph_2(\bzero)=\int p_2(\x_2)d^3x_2=\tr\rho_2=1$,
$\rho_1(\x,\x')$ remains unchanged on the diagonal $\x=\x'$.
This is because $\Hint$ is not changing the 
position of our our object particle, merely its momentum.
Since the mean position $\expec{\x_2}=\int p_2 \x_2 d^3x_2=\tr[\x_2\rho_2]=0$ 
vanishes (using \eq{r0defEq}), 
we have $\nabla\ph_2(\bzero)=\bzero$. In fact,
$|f|$ takes a maximum on the diagonal,
and the Riemann-Lebesgue Lemma shows that 
$|f|=|\ph_2|\le 1$ whenever $\x\ne \x'$, 
with equality only for the unphysical case
where $p_2$ is a delta function, \ie, where the location of the 
environment particle is perfectly known.
$\partial_i\partial_j |f(\bzero)| = -\M^\expec{\x_2\x_2^t}\M t^2/2\hbar^2$, so 
so the larger $\expec{\x_2\x_2^t}$ is
(\ie, the more spread out the environment particle is), 
the closer to the diagonal decoherence will suppress our density matrix.

Since $\M$ is the shear matrix of the force field $-\nabla V$,
we see that it is tidal forces that are causing the decoherence
--- the average force $\F$ simply contributes to the phase factor 
$e^{i\phi}$.
Specifically, the rate at which our object degrees of 
freedom $\r_1$ decohere grows with the tidal force that it 
exerts on the environment: if the environment particle is 
spread out with $\expec{\x_2\x_2^t}$ large, 
experiencing a wide range of forces from the object, 
object decoherence is rapid.
In the opposite situation, where the object is spread out and the
environment is not, the object will 
experience strong classical tidal forces but no decoherence.

\subsection{Relation between long-range and short-range decoherence}

Above we derived the effect of decoherence from long-range tidal forces.
Another interesting case that has been solved analytically \cite{JZ85}
is that of short-range interactions that can be modeled as
scattering events.
If the scattering takes place during short enough a time interval
that we can neglect the internal dynamics of the object,
then its reduced density matrix changes as \cite{collapse}
\beq{ScatteringEq}
\rho_1(\r,\r') \mapsto \rho_1(\r,\r')\ph\left({\r'-\r\over\hbar}\right),
\eeq
where $p(\q)$ is the probability distribution for 
the momentum transfer $\q$ in the collision.
This equation generalizes the scattering result of 
\cite{JZ85} by including the
effect of recoil.
The larger the uncertainty in momentum transfer, the stronger
the decoherence effect becomes, since 
widening $p$ narrows its Fourier transform $\ph$.
Changing the mean momentum transfer $\expec{\q}$
does not affect the decoherence, merely 
contributes a phase factor just as $\F$ did above.
Typically, the last factor in \eq{ScatteringEq}
destroys coherence down to scales of order the de Broigle
wavelength of the scatterer, 
with directional modulations from the angular dependence
of the scattering cross section.
Generalization to a steady 
flux of scattering particles
\cite{collapse} gives \eq{fEq1}.

\Eq{ScatteringEq} 
has striking similarities with the tidal force result
of \eq{JZfEq}: in both cases, the density matrix
gets multiplied by the Fourier transform of a
probability distribution.
If fact, up to uninteresting phase factors, 
we can rewrite our \eq{JZfEq} in exactly the form
of \eq{ScatteringEq} by redefining $p$ to be the 
probability distribution for momentum transfer 
$\q=\M(\x_2-\x_1)t$ due to tidal forces for a fixed $\x_1$, \ie,
\beq{pRedefEq}
p(\q)\equiv p_2(\x_2){d^3 x_2\over d^3 q}
={ p_2(\x_1 + \M^{-1}\q/t)\over t^3\det\M}.
\eeq
Fourier transforming this expression and substituting the
result into \eq{ScatteringEq}, we recover 
\eq{JZfEq} up to a phase factor.

Perhaps the simplest way to understand all these results is 
in terms of Wigner functions 
\cite{WignerFunc}.
If $W(\x_1,\p_1)$ is the Wigner phase space distribution
for the object particle, then any of the momentum-transferring
interactions that we have considered will 
take the form
\beq{Weq1}
W(\x_1,\p_1) \mapsto \int W(\x_1,\p_1-\q) p(\q,\x_1) d^3q
\eeq
for some probability distribution $p$ that may or may not 
depend on $\x_1$.
Since the density matrix
\beq{Weq2}
\rho_1(\x_1,\x_1') = 
\int W\left({\x_1+\x_1'\over 2},\p\right) e^{-i(\x-\x')\cdot\p}d^3 p
\eeq
is just the Wigner function Fourier transformed
in the momentum direction (and rotated by $45^\circ$),
the convolution with $p$ in \eq{Weq1} reduces to a simple multiplication
with $\ph$ in \eq{ScatteringEq}.





\begin{references}

\bibitem{Penrose89}
\rfbook\nn Penrose R;1989;The Emperor's New Mind;Oxford Univ. Press;Oxford

\bibitem{Penrose97}
\rfproc\nn Penrose R;1997;The Large, the Small and the Human Mind;
\nn Longair M;Cambridge Univ. Press;Cambridge

\bibitem{StappBook}
\rfbook\nnn Stapp H P;1993;Mind, Matter and Quantum Mechanics;
Springer;Berlin

\bibitem{Amit}
\rfbook\nnn Amit D J;1989;Modeling Brain Functions;Cambridge Univ. Press;Cambridge

\bibitem{Mezard}
\rfbook\nn M\'ezard M, \nn Parisi G\multiand\nn Virasoro M;1993;Spin Glass
Theory and Beyond;World Scientific;Singapore

\bibitem{Harvey}
\rfbook\nnn Harvey R L;1994;Neural Network 
Principles;Prentice Hall;{Englewood Cliffs}

\bibitem{Eeckman}
\rfbook\nnn Eeckman F H\dualand\nnn Bower J M;1993;Computation and 
Neural Systems;Kluwer;Boston

\bibitem{McMillen99}
\rf\nnn McMillen D R, 
\nnnn D'Eleuterio G M T\multiand\nnnn Halperin J R P;1999;Phys. Rev. E;59;6

\bibitem{Wigner62}
\rfproc\nnn Wigner E P;1962;The Scientist Speculates: 
an Anthology of Partly-Baked
Ideas, p284-302;\nnn Good I J;Heinemann;London

\bibitem{Wigner95}
\rfbook\nn Mehra J\dualand\nnn Wightman A S;1995;The Collected Works of E. P. Wigner, Vol. VI,
p271;
Springer;Berlin

\bibitem{Zeh70}
\rf\nnn Zeh H D;1970;Found. Phys.;1;69

\bibitem{Walker}
\rf\nnn Walker E H;1970;Mathematical Biosciences;7;131

\bibitem{Domash}
\rfproc\nnn Domash L H;1977;Scientific Research on TM;
\nnn Orme-Johnson D W\dualand\nnn Farrow J T;
Maharishi Univ. Press;{Weggis, Switzerland}

\bibitem{Stapp93}
\rf\nnn Stapp H P;1983;Phys. Rev. D;28;1386

\bibitem{Marshall}
\rf\nnn Marshall I N;1989;New Ideas in Psychology;7;73

\bibitem{Zohar}
\rfbook\nn Zohar D;1990;The Quantum Self;William Morrow;{New York}

\bibitem{Rosu}
\rf\nn Rosu H;1997;Metaphysical Review;3;{1, gr-qc/9409007}

\bibitem{Umezawa}
\rf\nnn Ricciardi L M\dualand\nn Umezawa H;1967;Kibernetik;4;44

\bibitem{Vitiello95}
\rf\nn Vitiello A;1996;Int. J. Mod. Phys.;B9;973-89


\bibitem{Hameroff}
\rf\nnn Hameroff S R\dualand\nnn Watt R C;1982;Journal of 
Theoretical Biology;98;549

\rfbook\nnn Hameroff S R;1987;Ultimate Computing: Biomolecular 
Consciousness and Nanotechnology;North-Holland;Amsterdam
 
\bibitem{Nano95}
\rfprep\nnn Nanopoulos D V;1995;hep-ph/9505374

\bibitem{Mavro95a}
\rfprep\nn Mavromatos N\dualand\nnn Nanopoulos D V;1995;hep-ph/9505401

\bibitem{Mavro95b}
\rfprep\nn Mavromatos N\dualand\nnn Nanopoulos D V;1995;quant-ph/9510003

\bibitem{Mavro95c}
\rfprep\nn Mavromatos N\dualand\nnn Nanopoulos D V;1995;quant-ph/9512021

\bibitem{Mavro98a}
\rf\nn Mavromatos N\dualand\nnn Nanopoulos D V;1998;Int. J. Mod. Phys B;12;{517, quant-ph/9708003}

\bibitem{Mavro98b}
\rfprep\nn Mavromatos N\dualand\nnn Nanopoulos D V;1998;quant-ph/9802063

\bibitem{Mavro99}
\rfprep\nn Mavromatos N;1999;J. Bioelectrochemistry \& Bioenergetics;48;273


\bibitem{Stapp91}
\rf\nnn Stapp H P;1991;Found. Phys.;21;1451


\bibitem{Zeh81}
\rn\nnn Zeh H D, quant-ph/9908084, 
{\it Epistemological Letters of the Ferdinand-Gonseth Association}
{\bf 63:0} (Biel, Switzerland, 1981)
  
\bibitem{Zurek91}
\rf\nnn Zurek W H;1991;Phys. Today;44 (10);36 

\bibitem{Scott96}
\rf\nn Scott A;1996;J. Consciousness Studies;6;484

\bibitem{Hawking97}
\rfproc\nn Hawking S;1997;The Large, the Small and the Human Mind;
\nn Longair M;Cambridge Univ. Press;Cambridge

\bibitem{Hepp99}
\rfproc\nn Hepp K;1999;Quantum Future;\nn Blanchard P\dualand\nn Jadczyk A;
Springer;Berlin

\bibitem{Neumann}
\rfbook\nn von~Neumann J;1932;Matematische Grundlagen der Quanten-Mechanik;Berlin;Springer

\bibitem{ZehTime}
\rfbook\nnn Zeh H D;1999;The Arrow of Time, 3rd ed.;Berlin;Springer


\bibitem{Zurek81}
\rf\nnn Zurek W H;1981;Phys. Rev. D;24;1516

\bibitem{Zurek82}
\rf\nnn Zurek W H;1982;Phys. Rev. D;26;1862

\bibitem{Zurek84}
\rfproc\nnn Zurek W H, reprint LAUR 84-2750;1984;Non-Equilibrium Statistical Physics;
\nn Moore G\dualand\nnn Sculy M O;Plenum;{New York}

\bibitem{Peres}
\rf\nn Peres E;1986;Am. J. Phys.;54;688

\bibitem{Pearle}
\rf\nn Pearle P;1989;Phys. Rev. A;39;2277

\bibitem{Gallis}
\rf\nnn Gallis M R\dualand\nnn Fleming G N;1989;Phys. Rev. A;42;38

\bibitem{Unruh}
\rf\nnn Unruh W H\dualand\nnn Zurek W H;1989;Phys. Rev. D;40;1071

\bibitem{Omnes97}
\rf\nn Omn\`es R;1997;Phys. Rev. A;56;3383

\bibitem{ZehBook}
\rfbook\nn Giulini D, \nn Joos E, \nn Kiefer C, \nn Kupsch J,
\nnn Stamatescu I O\multiand\nnn Zeh H D;1996;Decoherence and the Appearance
of a Classical World in Quantum Theory;Berlin;Springer

\bibitem{JZ85}
\rf\nn Joos E\dualand\nnn Zeh H D;1985;Z. Phys. B;59;223

\bibitem{collapse}
M. Tegmark, Found. Phys. Lett. {\bf 6}, 571 (1993).

\bibitem{Feynman72}
\rfbook\nnn Feynman R P;1972;Statistical Mechanics;Benjamin;Reading


 
\bibitem{Katz}
\rfbook\nn Katz B;1966;Nerve, Muscle, and Synapse;McGraw-Hill;{New York}

\bibitem{Schade}
\rfbook\nnn Schad\'e J P\dualand\nnn Ford D H;1973;Basic 
Neurology, 2nd ed.;Elsevier;Amsterdam

\bibitem{Cairns-Smith}
\rfbook\nnn Cairns-Smith A G;1996;Evolving the Mind;
Cambridge Univ. Press;Cambridge


\bibitem{Morell}
\rf\nn Morell P\dualand\nnn Norton W T;1980;Sci. Am.;242;74

\bibitem{Hirano}
\rfproc\nn Hirano A\dualand\nnn Llena J A;1995;The Axon;\nnn Waxman S G, 
\nnn Kocsis J D\multiand\nnn Stys P K;Oxford Univ. Press;{New York}

\bibitem{Ritchie}
\rfproc\nnn Ritchie J M;1995;The Axon;\nnn Waxman S G, 
\nnn Kocsis J D\multiand\nnn Stys P K;Oxford Univ. Press;{New York}

\bibitem{Wormholes}
\rf\nn Ellis J, \nn Mohanty S\multiand Nanopoulos D V;1989;Phys. Lett. B;221;113

\bibitem{Pearle76}
\rf\nn Pearle P;1976;Phys. Rev. D;13;857

\bibitem{GRW}
\rf\nnn Ghirardi G C, \nn Rimini A\multiand\nn Weber T;1986;Phys. Rev. D;34;470

\bibitem{Jackson}
\rfbook\nnn Jackson J D;1975;Classical Electrodynamics;Wiley;{New York}

\bibitem{ZHP}
\rf\nnn Zurek W H, \nn Habib S\multiand\nnn Paz J P;1993;Phys. Rev. Lett.;70;1187

\bibitem{gaussians}
\rf\nn Tegmark M\dualand\nnn Shapiro H S;1994;Phys. Rev. E;50;2538

\bibitem{Sataric}
\rf\nnn Satari\'c M V, 
\nnn Tuszy\'nski J A\multiand\nnn \v Zakula R B;1993;Phys. Rev. E;48;589

\bibitem{Rosu96a}
\rf\nnn Rosu H C;1997;Phys. Rev. E;55;2038

\bibitem{Rosu96b}
\rf\nnn Rosu H C;1998;Nuovo Cimento D;20;369

\bibitem{Stapp99}
\rn\nnn Stapp H S 1999,\\
{\it Attention, Intention, and Mind in Quantum Physics}
and {\it Quantum Ontology and Mind-Matter Synthesis},
available at {\it www-physics.lbl.gov/~stapp/stappfiles.html}.

\bibitem{Everett}
\rf\nn {Everett III} H;1957;Rev. Mod. Phys.;29;454

\rfbook\nn {Everett III} H;1986;The Many-Worlds Interpretation of
Quantum Mechanics, \nnn DeWitt B S\dualand\nn Graham N; 
Princeton Univ. Press;Princeton

\bibitem{MoreMWI}

\rf\nnn Wheeler J A;1957;Rev. Mod. Phys.;29;463;1957

\rf\nnn Cooper L M\dualand\nn {van Vechten} D;1969;Am. J. Phys;37;1212

\rf\nnn DeWitt B S;1971;Phys. Today;23;30

 
\bibitem{Lockwood}
\rfbook\nn Lockwood M;1989;Mind, Brain and the Quantum;
Blackwell;Cambridge

\bibitem{Deutsch}
D. Deutsch {\it The Fabric of Reality} (Allen Lane, New York, 1997).

\bibitem{Page95}
\rfprep\nnn Page D N A;1995;gr-qc/9507025

\bibitem{Donald97}
\rfprep\nnn Donald M J;1997;quant-ph/9703008

\bibitem{Donald99}
\rfprep\nnn Donald M J;1999;quant-ph/9904001

\bibitem{Vaidman}
\rn\nn Vaidman L 1996, quant-ph/9609006, 
{\frenchspacing\it Int. Stud. Phil. Sci.}, in press

\bibitem{Sakaguchi}
\rfprep\nn Sakaguchi T;1997;quant-ph/9704039

\bibitem{everett}
\rf\nn Tegmark M, quant-ph/9709032;1997;Fortschr. Phys.;46;855

\bibitem{TOE}
\rf\nn Tegmark M, gr-qc/9704009;1998;Annals of Physics;270;1




\bibitem{JamesBook}
\rfbook\nn James W;1890;The Principles of Psychology;Holt;{New York}

\bibitem{James}
\rfproc\nn James W 1904;1977;The Writings of William James, pp169-183;
\nnn McDermott J J;Univ. Chicago Press;Chicago


\bibitem{Eddington}
\rfbook\nn Eddington A;1920;Space, 
Time \& Gravitation; Cambridge Univ. Press;Cambridge

\bibitem{Lisewski}
\rfprep\nnn Lisewski A M;1999;quant-ph/9907052



\bibitem{WignerFunc}

\rf\nnn Wigner E P;1932;Phys. Rev.;40;749

\rf\nn Hillery M, \nnn O'Connell R H, 
\nnn Scully M O \& Wigner E P;1984;Phys. Rep.;106;121
 
\rfbook\nnn Kim Y S\dualand\nnn Noz M E;1991;Phase Space Picture of Quantum 
Mechanics: Group Theoretical Approach;World Scientific;Singapore

 
\end{references}
\end{document}